\documentclass[
 reprint,
 superscriptaddress,
 amsmath,amssymb,
 aps,
 pra,
showkeys
]{revtex4-2}

\usepackage{float}
\usepackage{multibib}
\usepackage{hyperref}
\usepackage{xcolor}
\hypersetup{
    colorlinks,
    linkcolor={blue!50!blue},
    citecolor={blue!50!blue},
    urlcolor={blue!80!black}
}
\usepackage{xcolor}
\usepackage{lmodern}
\usepackage{graphicx}% Include figure files
\usepackage{dcolumn}% Align table columns on decimal point

\begin{document}

\title{Comparative study of quantum emitter fabrication in wide bandgap materials using localized electron irradiation}

\author{Anand Kumar}
\email{anand.kumar@tum.de}
\affiliation{Department of Computer Engineering, School of Computation, Information and Technology, Technical University of Munich, 80333 Munich, Germany}
\affiliation{Abbe Center of Photonics, Institute of Applied Physics, Friedrich Schiller University Jena, 07745 Jena, Germany}
\author{Chanaprom Cholsuk}
\affiliation{Department of Computer Engineering, School of Computation, Information and Technology, Technical University of Munich, 80333 Munich, Germany}
\affiliation{Abbe Center of Photonics, Institute of Applied Physics, Friedrich Schiller University Jena, 07745 Jena, Germany}
\author{Mohammad N. Mishuk}
\affiliation{Department of Computer Engineering, School of Computation, Information and Technology, Technical University of Munich, 80333 Munich, Germany}
\affiliation{Abbe Center of Photonics, Institute of Applied Physics, Friedrich Schiller University Jena, 07745 Jena, Germany}
\author{Mouli Hazra}
\affiliation{Department of Computer Engineering, School of Computation, Information and Technology, Technical University of Munich, 80333 Munich, Germany}
\affiliation{Abbe Center of Photonics, Institute of Applied Physics, Friedrich Schiller University Jena, 07745 Jena, Germany}
\author{Clotilde Pillot}
\affiliation{Abbe Center of Photonics, Institute of Applied Physics, Friedrich Schiller University Jena, 07745 Jena, Germany}
\author{Tjorben Matthes}
\affiliation{Department of Computer Engineering, School of Computation, Information and Technology, Technical University of Munich, 80333 Munich, Germany}
\affiliation{Abbe Center of Photonics, Institute of Applied Physics, Friedrich Schiller University Jena, 07745 Jena, Germany}
\author{Tanveer A. Shaik}
\affiliation{Institute of Physical Chemistry and Abbe Center of Photonics (IPC), Friedrich Schiller University Jena, 07743 Jena, Germany}
\affiliation{Leibniz Institute of Photonic Technology (IPHT), 07745 Jena, Germany}
\author{Asl{\i} \surname{\c{C}akan}}
\affiliation{Department of Computer Engineering, School of Computation, Information and Technology, Technical University of Munich, 80333 Munich, Germany}
\author{Volker Deckert}
\affiliation{Institute of Physical Chemistry and Abbe Center of Photonics (IPC), Friedrich Schiller University Jena, 07743 Jena, Germany}
\affiliation{Leibniz Institute of Photonic Technology (IPHT), 07745 Jena, Germany}
\author{Sujin Suwanna}
\affiliation{Optical and Quantum Physics Laboratory, Department of Physics, Faculty of Science, Mahidol University, 10400 Bangkok, Thailand}
\author{Tobias Vogl}
\email{tobias.vogl@tum.de}
\affiliation{Department of Computer Engineering, School of Computation, Information and Technology, Technical University of Munich, 80333 Munich, Germany}
\affiliation{Abbe Center of Photonics, Institute of Applied Physics, Friedrich Schiller University Jena, 07745 Jena, Germany}

\date{\today}% It is always \today, today,
             %  but any date may be explicitly specified
             
\begin{abstract}
Quantum light sources are crucial foundational components for various quantum technology applications. With the rapid development of quantum technology, there has been a growing demand for materials with the capability of hosting quantum emitters. One such material platform uses fluorescent defects in hexagonal boron nitride (hBN) that can host deep sub-levels within the band gap. The localized electron irradiation has shown its effectiveness in generating deep sub-levels to induce single emitters in hBN. The question is whether localized (electron beam) irradiation is a reliable tool for creating emitters in other wide bandgap materials and its uniqueness to hBN. Here, we investigate and compare the fabrication of quantum emitters in hBN and exfoliated muscovite mica flakes along with other 3D crystals, such as silicon carbide and gallium nitride, which are known to host quantum emitters. We use our primary fabrication technique of localized electron irradiation using a standard scanning electron microscope. To complement our experimental work, we employ density functional theory simulations to study the atomic structures of defects in mica. While our fabrication technique allows to create hBN quantum emitters with a high yield and high single photon purity, it is unable to fabricate single emitters in the other solid-state crystals under investigation. This allows us to draw conclusions on the emitter fabrication mechanism in hBN, which could rely on activating pre-existing defects by charge state manipulation. Therefore, we provide an essential step towards the identification of hBN emitters and their formation process.
\end{abstract}


\maketitle

%\tableofcontents
\section{Introduction}
\section{Introduction}
Quantum emitters in solid-state crystals have garnered considerable attention, driven by the rapid advancement of quantum technology applications such as quantum computing, quantum communication, and quantum sensing  \cite{Degen2017, Kanae, abasifard2023ideal, ali, Samaner2022}. The discovery of quantum emitters based on defects in wide bandgap materials has significantly advanced this field \cite{Aharonovich2016, Tran2016, Senichev2021, Lohrmann2015, Wang2018, Stachurski2022}. Quantum emitters have been used in a wide variety of applications, most prominently in magnetometry and imaging \cite{Exarhos2019, Huang2022}, but also in quantum key distribution \cite{ali, Samaner2022}, fundamental quantum physics tests \cite{Vogl2021}, thermometry \cite{Chen2020}, pressure sensing \cite{andreas}, quantum computing \cite{Conlon2023}, quantum memories \cite{Pfaff2013,nos-qmemory,takla-qmemory}, and as nodes in a quantum network \cite{Ruf2021}.\\
\indent Probably the most-well studied solid-state quantum emitter is the nitrogen vacancy center in diamond \cite{Doherty2013} and related defects, such as the group-IV color centers \cite{Bradac2019}. Quantum emitters in two-dimensional materials such as semiconducting transition metal dichalcogenides (TMDs) \cite{carmem, Klein2019} and insulating hexagonal boron nitride (hBN) \cite{Aharonovich2016, Tran2016} offer the advantage of intrinsically high photon out-coupling, as a defect in an atomically thin material is not surrounded by any high refractive index and therefore not limited by any total internal or Fresnel reflection \cite{Vogl2019}. In addition, 2D materials can be easily attached to photonic components through van der Waals forces, making them outstanding candidates for integrated photonics and waveguides platforms \cite{Vogl2017}.\\
\indent The quantum emitters can be fabricated using various methods such as strain activation using nanostructure \cite{carmem}, mechanical damage using the tip of an atomic force microscope (AFM) \cite{Xiaohui2021}, thermal activation of naturally occurring defects \cite{Tran2016}, plasma \cite{Vogl2018} and chemical etching \cite{Chejanovsky2016}, as well as energetic irradiation \cite{Klein2019, 10.1038/s41467-019-09219-5,tran2016robust, sumin2016engineering, Fournier2021, Gale2022, Kumar2023}. Of particular interest is the formation of emitters at pre-defined locations, while minimizing the impact of the crystal environment, for which localized irradiation has shown to be useful. The response of materials to the irradiation by energetic particles such as electrons and ions has been studied intensively \cite{2ddefect}. Irradiation can create new defects in solids by knocking out atoms or implanting impurities, but also modify or activate existing defects by breaking and re-organizing bonds or changing the charge state. This can ultimately result in the modification of their physical and chemical properties. Recent works have shown the high yield of quantum emitters in TMDs and hBN by modifying the surface properties or doping \cite{Gale2022, Fournier2021, Klein2019, Kumar2023} and arrays of photophysically identical emitters with correlated dipoles \cite{kumar2023polarization}, all using electron irradiation of hBN or a focused helium beam in case of MoS$_2$ \cite{Klein2019}.\\
\indent The emitter formation process through electron irradiation in hBN remains inadequately understood. The emitters not only differ in terms of formation depth when compared to alternative methods such as plasma-activated emitters followed by thermal annealing \cite{Vogl2019} but also exhibit varying photo-physical properties. Some present randomly oriented dipoles \cite{PhysRevApplied.18.064021}, while other shows correlated dipole direction \cite{kumar2023polarization}. Additionally, there are some emitting blue light \cite{Gale2022, Fournier2021}, and others having their optical transitions in the yellow regions \cite{Kumar2023,kumar2023polarization}.\\
\indent While our previous work \cite{Kumar2023} focused on localized irradiation, which resulted in yellow emitters with high yield, it is still unable to address the uniqueness of the electron irradiation process to hBN and its applicability to other wide band gap materials. This is because, upon irradiation, we notice the formation of the surface complex (black dots in SEM images), which could be responsible for generating the emitters. Based on this idea, here, we explore how and whether our fabrication method can promise to generate emitters independent of material choice. For this, we consider two different types of hBN crystals grown differently, and analyze whether this strategy is applicable to different wide band gap materials. Moreover, with this scenario, we make use of the significant conclusion discussed in \cite{Gale2022}, in which it has been demonstrated that the generation of blue emitters shows a strong correlation with a sharp UV emission at 305 nm attributed to the presence of carbon dimers.  \\
\indent More precisely, along with hBN, we considered mechanically exfoliated 2D muscovite mica with a band gap above 5 eV \cite{Frisenda2020, Kalita2016} and 3D crystals such as silicon carbide and gallium nitride. While mica is not studied very well from the prospective to host quantum emitters, both the other 3D crystals are well known to host optically active quantum emitters \cite{Lohrmann2015, Wang2018, Castelletto2013, Senichev2021, Stachurski2022, Meunier2023}. Furthermore, to understand the material growth sensitivity of electron irradiation, we utilize two different types of hBN crystals grown through different methods obtained from separate vendors. We first created an array of high-quality emitters in our standard hBN crystal and then replicated the same fabrication process on the other samples. In addition, we also try high-temperature annealing and check if any newly created defects require thermal activation. Our experiments and conclusions are supported using density functional theory (DFT) calculations, which reveal the electronic structure of potential defects in mica and allow us to infer whether the defect should be optically active.\\
\begin{figure*}
    \includegraphics[width = 1\textwidth]{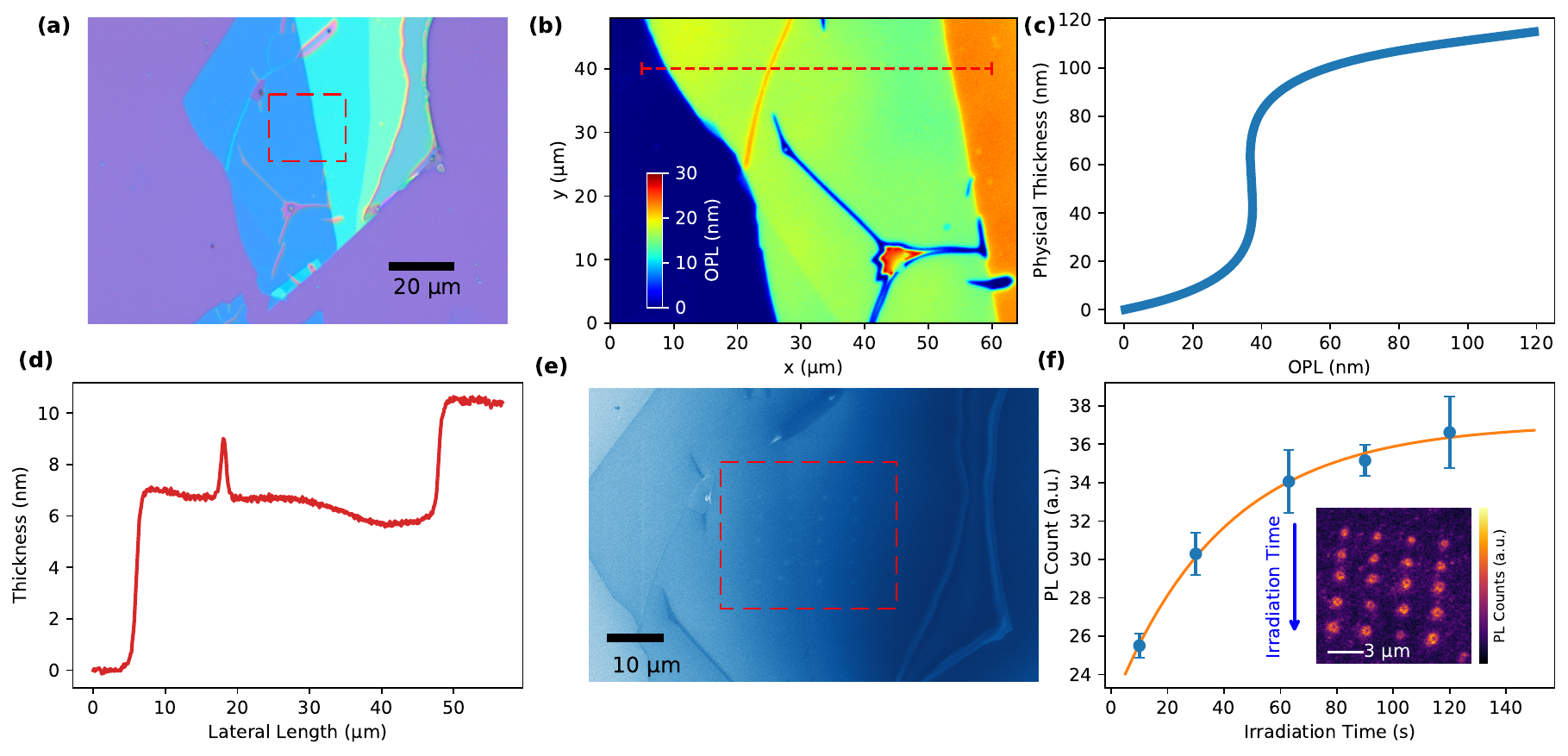}
    \caption{(a) The bright field optical microscope image of an exfoliated mulit-layer Mica flake on a standard Si/SiO$_2$ substrate with 290 nm of oxide thickness. (b) The height map of the the flake measured using phase shift interferometer (PSI). The color-bar present the optical path length (OPL) through the flake. (c) The RCWA simulation of the physical thickness as function of OPL of the mica flake, calibrated with AFM and PSI measurements. (d) The actual physical thickness of the flake along the dashed line and measured thickness is in the range of 6 nm to 10 nm. (e) SEM image of the flake after electron irradiation. The radiation sites are visible as distinctive dots within the dashed rectangle. (f) The averaged PL count versus electron beam irradiation time presenting exponential saturation behaviour. The average PL count extracted from the inset PL map created using 470 nm pulsed excitation laser. }
    \label{fig-1}
\end{figure*}

\section{Experimental details}
\subsection{Sample preparation}
To create hBN emitters, we first mechanically exfoliate from bulk crystal onto an organic polymer sheet of PDMS (polydimethylsiloxane) using scotch tape. Afterwards, a suitable thin flake is transferred onto a grid-patterned standard Si/SiO$_2$ substrate with a 290 nm thermal oxide layer by dry stamping (see Methods). We have used bulk hBN from HQ Graphene (named hBN-1 in the following) and 2D Semiconductors (hBN-2), which produce their crystals using different processes, e.g., 2D Semiconductors utilize a high-pressure anvil cell growth method. It is, therefore, likely that the samples have a different intrinsic doping of impurities.\\
\indent The muscovite mica flakes are exfoliated and transferred in the same fashion as mentioned above. Figure \ref{fig-1}(a) shows an optical image of a transferred mica flake. The thickness map is created using a phase-shift interferometer (PSI) that measures the optical path length (OPL) through the flake and is shown in Figure \ref{fig-1}(b). To determine the physical thickness of the OPL, we use Rigorous Coupled-Wave Analysis (RCWA) simulations. This method is similar to what has been previously done for TMDs \cite{Yang2016} and hBN \cite{Vogl2018}. The RCWA model depicted in Figure \ref{fig-1}(c) is calibrated using real thickness measurements obtained with an AFM (see Supplementary Section S1). Once the RCWA model is calibrated, the use of PSI proves to be both accurate and faster than the commonly employed AFM thickness measurements \cite{Yang2016, Vogl2018, Vogl2019}. It is important to note, however, that this technique can only yield accurate results for OPL thicknesses up to 30 nm (for mica on Si/SiO$_2$ substrate with a 290 nm thermal oxide layer), as the relation between physical and optical thickness follows an S-curve due to interference (see Figure \ref{fig-1}(c)). The physical flake thickness through the dashed line in the OPL map is shown in Figure \ref{fig-1}(d). The flake on which the subsequent electron irradiation is carried out has a thickness of around 6 nm (blue area in the microscope image) and around 10 nm (teal area in the microscope image).\\
\indent The 3D crystals, silicon carbide (4H) and gallium nitride (n-type doped on sapphire), were directly acquired from MSE Supplies and used as received.

\subsection{Electron irradiation}
We performed localized electron irradiation using a scanning electron microscope (SEM) to fabricate defect-based quantum emitters into the materials. This method has demonstrated its effectiveness in creating quantum emitters within hBN, as substantiated by prior studies \cite{Fournier2021, Gale2022, Kumar2023}. The high lateral resolution of SEM allows us to irradiate the sample within the limit of electron beam diameter. The irradiation is performed at a user selected spot within a pre-acquired SEM image (see Supplementary Section S2). Figure \ref{fig-1}(e) presents an SEM image of a multi-layer mica flake post-irradiation, clearly showcasing the irradiation spots. Here, each spot is irradiated using an accelerating voltage of 3 kV and an electron current of 25 pA with a dwell time of 10 s (see Methods). While the theoretical resolution limit of SEM in field-free mode is around 10 nm, the used electron beam settings and beam alignment result in an electron beam diameter of around 300 nm. Since this is already the diffraction limit of our optical setup in subsequent measurements, we did not perform any further beam alignments. This resulted in an electron fluence of $7.7 \times 10^{17}$ cm$^{-2}$ at each irradiated spot. Note that we used low electron fluence (10$^{13}$ cm$^{-2}$) to record the image of the flake to avoid the creation of randomly located emitters due to electron beam scanning during the imaging process.  \\
\indent To control the electron dose precisely, we adjusted the irradiation time during the irradiation process. Additionally, we performed the irradiation process with varying electron irradiation times on one of the flakes. The resulting PL map of the irradiated spot within the flake is displayed in the inset of Figure \ref{fig-1}(f). The average PL intensity saturates with the irradiation time as shown in Figure \ref{fig-1}(f), where average PL intensity is determined by averaging over the same number of pixels for each spot (see Supplementary Section S3). This behaviour suggests that defects in the mica flake are constrained by an upper limit, resembling the controlled creation of emitters in hBN through the irradiation process \cite{Roux2022, Gale2022, Kumar2023}. These findings underscore the precision and controllability of defect engineering within the mica flakes, promising exciting prospects for applications in quantum emitters fabrication.

\section{Results and discussions}
\subsection{Optical characterization}

\begin{figure*}
    \includegraphics[width = 1\textwidth]{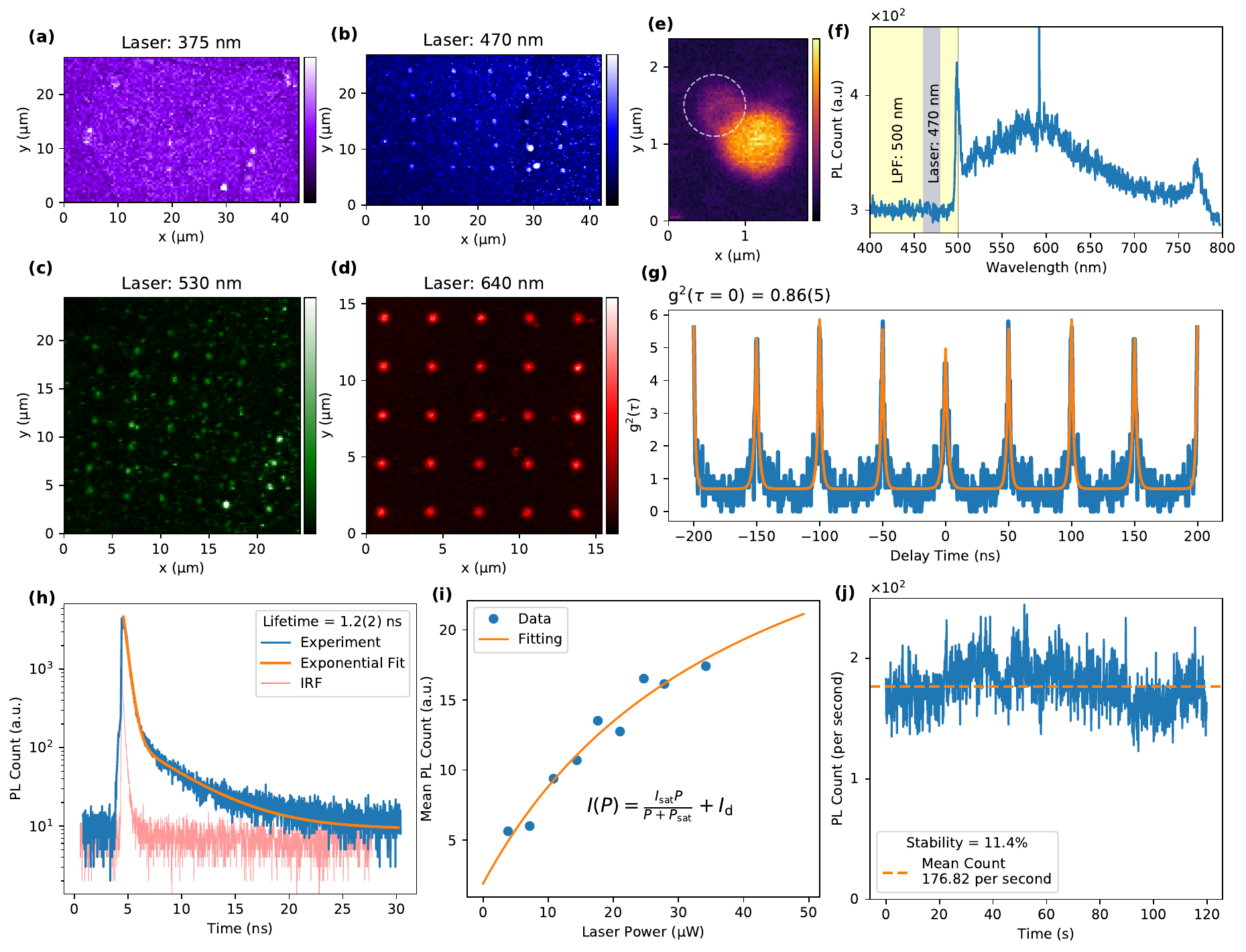}
    \caption{(a)-(d) PL map of a mica flake created using a pulsed excitation laser of wavelength (a) 375 nm (b) 470 nm (c) 530 nm and (d) 640 nm with a repetition rate of 20 MHz. The bright spots in the PL map correspond to the irradiated spot. (e) A zoomed in PL map of an irradiated spot under 470 nm pulsed excitation laser. A diffraction limited spot is is indicated close to the irradiated spot. (f) The measured spectrum from the irradiated spot presenting a broad peak around 600 nm. (g) The second-order correlation measurement with pulsed excitation laser at a wavelength of 470 nm with repetition rate of 20 MHz. A model function is used to fit the experimental curve and extracted g$^{(2)}(0)$ is $0.86 \pm 0.05$. (h) A typical lifetime decay of at one of the bright spot revealing the typical lifetime around 1.2 ns under 470 nm excitation laser. (i) The power dependent PL showing a saturation curve similar to the two level system. The saturation intensity of 35.18 counts per second and saturation power 40.75 $\mu$W are extracted from the model function as indicated. (j) The time trace of an emitter presenting a stable emission. The mean and standard deviation of the count rate are 175.82 Hz and 20.09 Hz, implying a relative stability of 11.4\%.  }
    \label{fig-2}
\end{figure*}

\begin{figure*}
    \includegraphics[width = 1\textwidth]{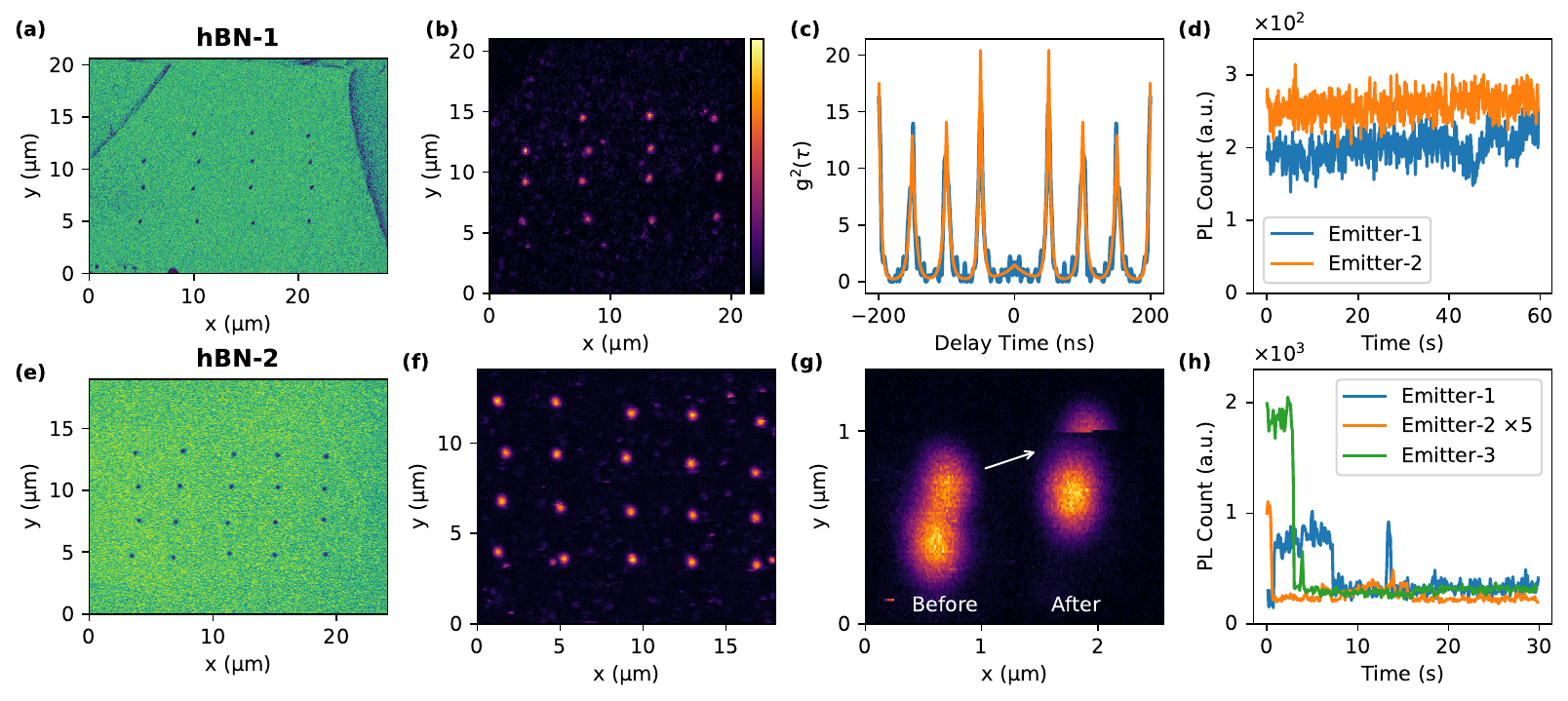}
    \caption{Top row (sample hBN-1): (a) The resulting (false-color) SEM image after localized irradiation of the flake, showcasing the distinct irradiated regions. (b) The PL map of the flake using 530 nm pulsed excitation laser resulting bright emission from the irradiated spots. (c) The second-order correlation measurement from one of the irradiated spot. A model function is used to fit the experimental curve and extracted g$^{(2)}(0)$ is 0.084(2). (d) PL count versus time from the emitters revealing a stable PL emission of two emitters. Bottom row (sample hBN-2): (e) The (false-color) SEM images of an irradiated flake. (f) The resulting PL map of the irradiated flake presenting bright emission from the irradiated spot. (g) Two consecutive zoomed-in PL maps of the same irradiated spot, providing visual evidence of emitter bleaching between the before and after PL maps. (d) The PL count rate versus time presenting the unstable emission (bleaching and blinking) from various emitters under 530 nm pulsed excitation laser in sample hBN-2.}
    \label{fig-3}
\end{figure*}

To investigate our fabrication methodology of quantum emitters, we optically characterized the samples using a commercial fluorescence lifetime imaging microscope (see Methods). We first investigated irradiated mica flakes. We performed all the measurements at room temperature using different pulsed excitation lasers with a repetition rate of 20 MHz. We always excite below bandgap to exclusively excite any states that were created during the irradiation process and at the same time to avoid the excitation of the free exciton across the bandgap. Figures \ref{fig-2}(a)-(d) present the PL maps of the irradiated mica flakes using 375 nm, 470 nm, 530 nm and 640 nm pulsed excitation lasers, respectively. Under 375 nm excitation laser, we observe weak emission from the irradiated spots indicating very inefficient excitation in the UV region. However, we notice bright emission from the irradiated spots under different excitation lasers as shown in Figure \ref{fig-2}(b)-(d). Bright emission in the PL map corresponds to the irradiated spots as can be confirmed by SEM and PL images. We also did not observe any correlation in between bright emission from irradiated spots and the flake thickness.\\
\indent Furthermore, we see diffraction-limited isolated bright spots in the close vicinity of the irradiated spots, as shown in Figure \ref{fig-1}(e), when a zoomed-in PL map is created using a 470 nm pulsed excitation laser. This is similar to the previously studied hBN irradiation experiments, which were responsible for the generation of blue and yellow single photon emitters around the irradiated spots \cite{Fournier2021, Gale2022, Kumar2023, kumar2023polarization}. However, we did not find any sharp spectral feature resembling ZPL, as recorded using a 470 nm excitation laser in combination with a long-pass filter with a cut-off wavelength at 500 nm to suppress the excitation laser. We instead observe a broad peak centred around 600 nm as shown in Figure \ref{fig-2}(f). The emission spectrum is similar to that observed from the irradiated spots in hBN \cite{Kumar2023, kumar2023polarization}. This observation could indicate that our irradiation process may indeed lead to the formation of similar defect complexes in mica, possibly induced by electron irradiation and potentially associated with deposited carbon. Note that we extended the exposure time to acquire the spectrum effectively due to the relatively weak PL emission. As a result of this longer exposure time, we also captured certain experimental artefact peaks in our measurements. Notably, these artefact peaks were also present in our background spectrum measurements, which were taken at a location away from the irradiated spot and subjected to the same extended exposure time (see Supplementary Section S4).\\
\indent To further investigate the quantum nature of emission, we measure the second-order correlation function (g$^{(2)}$) from the various irradiated spots. In Figure \ref{fig-2}(e), we present an experimentally measured g$^{(2)}$ curve, acquired using a 470 nm pulsed excitation laser with a repetition rate of 20 MHz. To analyze the data, we employed a model function to fit the g$^{(2)}$ curve and extracted a g$^{(2)}(0)$ around 0.86(5), which is larger than 0.5 (the generally accepted criterion for single photon emission due to a non-zero overlap with the single photon Fock state). Unlike previous recent studies which introduced organic fluorescent molecules from solvents \cite{Neumann2023}, we have not been able to create single emitters in mica using our electron irradiation process. The excited state lifetime of the emitters is estimated to be around 1.2 ns. The lifetime decay curve is fitted using a bi-exponential model function, as shown in Figure \ref{fig-2}(h). It is worth noting that our fitting routine accounts for the instrument response function (IRF) shown in Figure \ref{fig-2}(h), which has a relatively faster response time below 100 ps. Additionally, we examined the intensity from the irradiated spot as a function of excitation laser power, revealing the typical saturation curve for the two-level system, as depicted in Figure \ref{fig-2}(i). We fitted the experimental data using a model function revealing the average saturation power to be around 40.75 $\mu$W, and saturation intensity is 35.18 counts per second. The observed saturation count rate is significantly lower compared to the typical hBN emitters, including those induced by irradiation \cite{Fournier2021, Gale2022, Kumar2023}. This implies that mica irradiated emitters have poor quantum efficiency, suggesting that mica may not be a suitable material for generating emitters through the irradiation process. Remarkably, our experimental observation did not reveal any photo-bleaching or blinking effects under pulsed excitation. We measured the photon count rate over an extended period as shown in Figure \ref{fig-2}(j), which revealed a stable emission with a stability of 11.4\%. \\  
\indent To gain deeper insights into the generality of our fabrication methodology in generating quantum emitters, we extended our irradiation experiment on various other insulating materials, including hBN-1, hBN-2, silicon carbide (4H-SiC) and gallium nitride (GaN) crystals. For hBN-1, we initially exfoliated a multi-layer flake onto a standard Si/SiO$_2$ substrate and conducted localized irradiation at a chosen spot within the flake, employing a dwell time of 10 seconds. This resulted in a similar electron fluence as in the case of mica irradiation (see Methods). Figure \ref{fig-3}(a) shows the resulting SEM image of the flake. Subsequently, we recorded the PL map of the irradiated flake using a 530 nm pulsed excitation laser as shown in Figure \ref{fig-3}(b). The bright spot in the PL map corresponds to the irradiated spots, as evident in the SEM image. This is consistent with our previous study that yields single emitters at 575 nm \cite{Kumar2023}. To further understand the quantum nature of emission, we performed g$^{(2)}$ measurements using a 530 nm pulsed excitation laser with a repetition rate of 20 MHz. Figure \ref{fig-3}(c), reveal a clear g$^{(2)}$ dip indicating quantum nature of emission. The g$^{(2)}$(0) value of 0.084(2) is extracted from the model fitting function. We also noticed a stable emission from the quantum emitters with no instances of blinking and photo-bleaching, as observed in Figure \ref{fig-3}(d). The further photo-physical properties of as-fabricated emitters are also summarized in our previous works \cite{Kumar2023, kumar2023polarization}.\\
\indent For hBN-2, we conducted irradiation using the same electron beam parameters (see Methods). The resulting SEM image of the flake is shown in the \ref{fig-3}(e), and the corresponding PL map of the flake is presented in Figure \ref{fig-3}(f), revealing bright emission from the irradiated spots. The PL map is generated using a 530 nm pulsed excitation laser with a similar average laser power (see Methods). The bright emission indicates the generation of a similar defect complex due to localized irradiation as in hBN-1. A zoomed-in PL map around the irradiated spot reveals the isolated diffraction-limited bright spot identical to what is observed in hBN-1, which presents stable quantum emission. However, in the subsequent PL map, it is evident that the emitter eventually bleached out, as shown in Figure \ref{fig-3}(g). Furthermore, investigation reveals PL count drops to the detector noise level within 10 s under pulse excitation, as shown in Figure \ref{fig-3}(h). This instability is observed in almost every studied emitter in hBN-2 crystal (see Supplementary Section S5). This could be attributed to the unstable charge carriers in the excited state, potentially due to the material's intrinsic doping, which depends on the growth conditions.\\
\indent In addition to the hBN, our irradiation experiment on SiC (4H-semi insulating) and GaN do not activate any emitters or emitters ensembles as evident by PL map (see Supplementary Section S6). We found that this is independent of the electron beam dose and electron accelerating voltage. We can draw already two conclusions from our experiments: (i) the electron irradiation process works very well to generate single photon emitters in hBN but not in other wide bandgap materials. This indicates that the hBN defects are likely lattice defects and not deposited complexes on the surface, at least using this fabrication method. If latter was the case, this should also work on other materials as a substrate. And (ii), while we could create single photon emitters in both hBN samples, there seems to be a different intrinsic defect density that instabilizes the defects in hBN-2.

\subsection{Theoretical calculations}
\begin{figure*}
    \includegraphics[width = 1\textwidth]{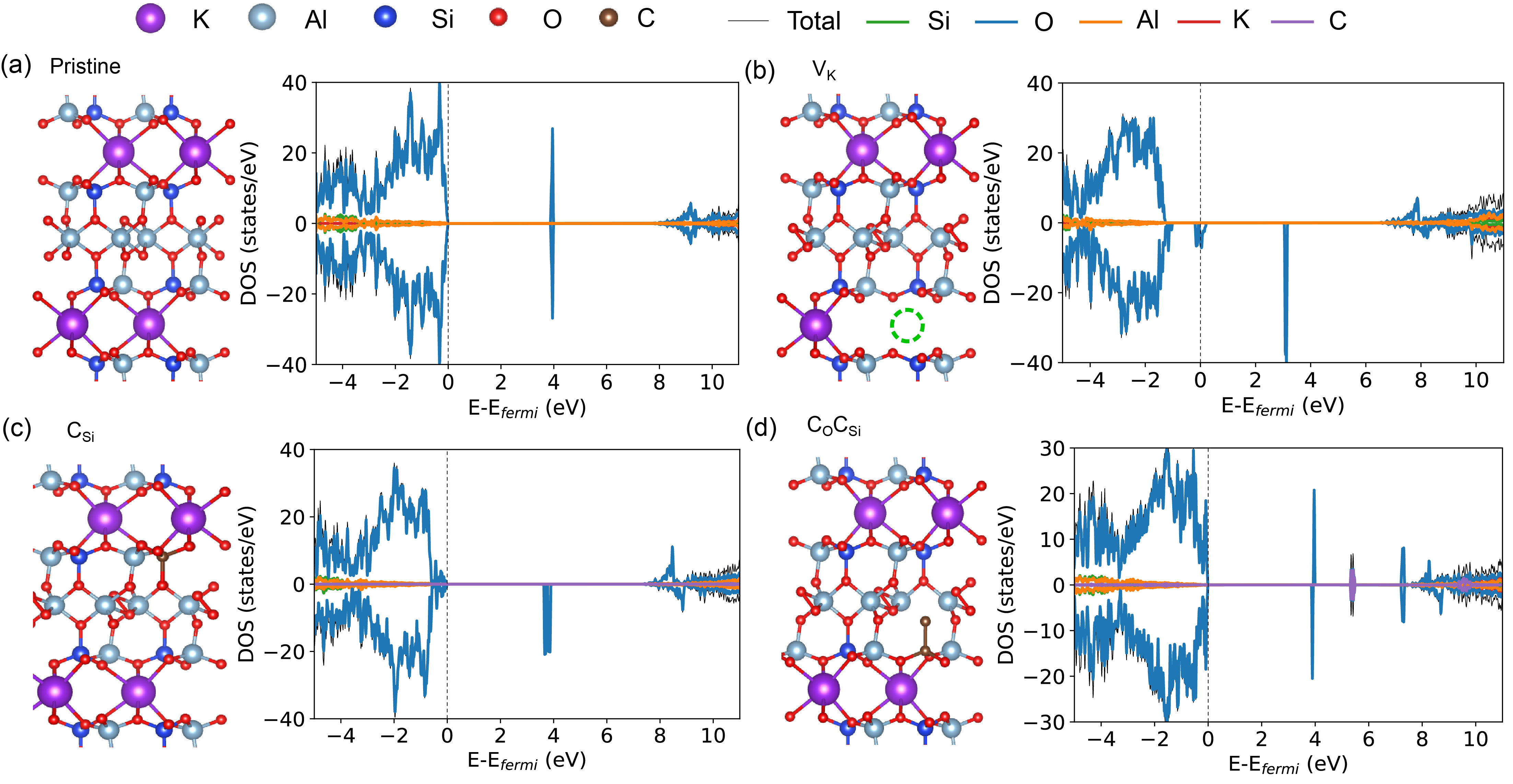}
    \caption{Lattice structures and density of states of (a) a pristine and three defects consisting of (b) a native vacancy, (c) an antisite with carbon doping, and (d) defect complexes.}
    \label{fig:band}
\end{figure*}
It is known that hBN hosts a large variety of optically active emitters \cite{Cholsuk2022}. The question arises if this is true for mica as well. Here, we would determine if our fabrication method does not work on mica or if there are generally not many optically active defects using DFT calculations (See Methods for the computational details). Here, to estimate the optical signatures, we consider the spin polarization as it should be preserved for the allowed optical transition. Figure \ref{fig:band}(a) depicts the electronic structure of the pristine mica with a bandgap of 7.91 eV. Due to this wide bandgap property, mica should, in principle, support defect levels similar to other wide bandgap materials like hBN  \cite{Cholsuk2022, Cholsuk2023, Sajid2018} and diamond \cite{Gali2019, Gupta2019, Seo2017}. In these materials, single photon emission originates from a transition between two-level states of occupied and unoccupied defect states, unlike TMDs that are some time from the transition between strained valence and conduction bands \cite{Parto2021, Linhart2019}. Consequently, we calculate 20 different point defects, primarily native and antisite defects but also carbon impurities, since an SEM typically bonds carbon to the surface \cite{Kumar2023}. For this, we aimed to investigate the existence of two-level states localized far away from the valence and conduction bands greater than 1 eV, the so-called deep-lying defect states.\\
\indent In Figure \ref{fig:band}(b), the V$_\text{K}$ defect introduces two-level states between occupied and unoccupied states contributed by oxygen atoms. However, these are shallow levels which have full occupation at room temperature. Considering a carbon dopant at different sites in Figures \ref{fig:band}(c) and (d), both cases do not exhibit any two-level states. Although it likely has an unoccupied state at 5 eV, which can accept some excited electrons, this large transition energy is out of reach for our experiments. In addition, we expect that this system will have a low transition rate. Together with the absence of an occupied isolated defect state, both are unlikely to act as a single photon emitter. It is worth noting that all defects seem not to localize at both bottom-most and top-most states. This indicates that the defect in mica has little influence on deep-lying two-level states. All these behaviours have also been observed in the rest of the defects shown in Supplementary S7 except for Al$_\text{O}$, where its desired two-level states exist. Thus, this implies that mica hosts many optically inactive defects compared to hBN \cite{Cholsuk2022, Sajid2018}. As a result, it can be concluded that although mica offers a wide bandgap, which in principle can host many defects, the electronic states of those doped defects do not appear in the bandgap region but rather around the valence and conduction bands. In other words, mica is not feasible to support single photon emitters, at least not as suitable as hBN.

\section{Conclusion}
We extensively demonstrate the investigation of quantum emitter fabrication using a localized electron irradiation process and provided valuable insights into the effectiveness of electron irradiation across various other insulating materials. Our findings reveal that localized electron irradiation effectively generates single photon emitters in hBN with an optical transition in the yellow region. However, we observe the photo-stability issue between hBN crystals grown differently, attributing to the different intrinsic defect densities or doping levels. This indicates that emitter generation using electron irradiation relies on activation or modification of the charge state of pre-existing defects.\\
\indent By testing this process on other materials, we can distinguish between a surface complex that is deposited onto the material during our electron irradiation and such activation of lattice defects. This is in particular important as a recent study using organic solvents has found that (at least some) of the hBN as well as mica emitters to be related to organic molecules adhered to the surface of hBN and mica, respectively \cite{Neumann2023}. We can rule out this possibility for our hBN defects. In addition, we have found that our method does not activate pre-existing defects in the other investigated materials. This could have several reasons where we can only speculate. Maybe there are simply no intrinsic defects that can be activated with an electron beam. Of course, the electrons do not have enough energy to knock out atoms and produce, e.g., vacancies.\\
\indent Our experimental evidences on mica are also supported by theoretical DFT simulation. The theoretical findings on mica indicate that it is unlikely to support single photon emitters due to the lack of two-level defect states within its band gap. Therefore, the bright emission observed in the irradiated mica sample may be attributed to the generation of surface complexes, similar to a previous observation \cite{Neumann2023}.\\
\indent Overall, our findings highlight the intricate relationship between the choice of insulating material and the effectiveness of localized irradiation in creating quantum emitters. While this methodology has proven robust in some materials like hBN, its efficacy varies. It presents exciting possibilities and challenges in the quest to engineer quantum emitters for diverse applications in quantum technologies and photonics. Further research and optimization efforts are warranted to harness the full potential of this fabrication technique across different material platforms.

\section*{Methods}
\subsection*{Sample preparations} 
We used a standard Si/SiO$_2$ substrate (purchased from Microchem) with a 290 nm thermally grown oxide layer. Afterwards, the substrate is processed using electron beam lithography and a metal lift-off process to realize a cross-grid pattern. The cross gird patterned is used to navigate to the target flake during sample fabrication and characterization processes. The bulk crystal for exfoliation is commercially purchased from hq Graphene (Muscovite mica and hBN-1), 2D Semiconductor (hBN-2). Thin mica, hBN-1 and hBN-2 flakes were first mechanically exfoliated onto a polymer sheet (PDMS, Polydimethylsiloxane) using scotch tape. Afterwards a suitable thin flake is selected under bright field optical microscope and then transferred to a cross grid patterned standard Si/SiO$_2$ substrate. Additionally, we acquired a 10$\times$10 mm$^2$ 4H semi-insulating-type silicon carbide crystal substrate from MSE Supplies. In addition, we obtained a 2-inch undoped n-type gallium nitride template, which was grown on a sapphire (0001) substrate.

\subsection*{Electron irradiation}
The fabrication of electron-irradiated emitters was conducted using a Helios NanoLab G3. To ensure precise electron beam alignment without unintended irradiation, this alignment procedure was carried out at a separate location on the substrate close to the target flake. For sample navigation and image acquisition of the flakes, we employed a low electron fluence of $1.4\times 10^{13}$ cm$^{-2}$, with an accelerating voltage of 3 kV and a electron current of 25 pA. To fabricate the emitters, each chosen spot is irradiated with dwell time of 10 s with fluence value of $7.7\times 10^{17}$ cm$^{-2}$ at a voltage of 3 kV and current of 25 pA. 

\subsection*{Optical characterization}
The optical investigation of the quantum emitters was conducted using a commercial time-resolved confocal microscope (PicoQuant MicroTime 200). This setup offers four linearly-polarized excitation lasers with wavelengths of 375 nm, 470 nm, 530 nm, and 640 nm, each with pulse lengths ranging from 40 to 90 ps (FWHM), depending on the wavelength. The excitation power used for all measurements was maintained at approximately 30 $\mu$W, unless specified otherwise. To effectively distinguish the emitted signals from the excitation laser, the detection path of the setup was equipped with a variety of notch filters, including both long-pass and bandpass filters (inserted depending on the specific laser wavelength). Moreover, the setup featured a dual configuration of single photon avalanche diodes (SPADs) positioned in both arms of a 50:50 beam-splitter. This configuration enabled us to measure the second-order correlation function (g$^{(2)}$) and conduct other relevant photon statistics analyses. For the acquisition of PL maps for each individual flake, the stage was meticulously scanned with a dwell time of 5 ms, and the laser operated at a repetition rate of 20 MHz. The PL signal is collected using a 100$\times$ dry immersion objective with a high numerical aperture (NA) of 0.9. The data analysis of the correlation function as well as lifetime measurements is performed with the built-in software (that also takes the instrument response function into account). The data acquisition times for these measurements was  1 min per emitter. The spectrometer (Andor Kymera 328i) is attached to one of the exit ports of the optical setup to collect the spectrum of the emitters.

\subsection*{DFT calculations}
The DFT calculations were carried out using the plane-wave Vienna Ab initio Simulation Package (VASP) \cite{vasp1,vasp2}. A projector augmented wave (PAW) was employed for treating the core nuclei and valence electrons \cite{paw,paw2}. A crystal structure of bulk pristine muscovite mica (KAl$_3$Si$_3$O$_{10}$[OH]$_2$) containing 76 atoms from the database \cite{mica_structure} was fully optimized with a cutoff energy of 450 eV and the convergence threshold at 10$^{-4}$ eV until the convergence force is less than 0.02 eV/\AA. A 5$\times$4$\times$3 Monkhorst–Pack reciprocal space grid was implemented for Brillouin zone integration \cite{kpoint}. After structural relaxation of the pristine structure, we obtained the lattice parameters at $a = 5.19$ \AA, $b = 9.00$ \AA, $c = 20.10$ \AA, and $\beta = 95.18^\circ$, corresponding to other calculations and experiments \cite{Franceschi2023,Mukherjee2022,Wu2022}. It is important to note that experimental band gaps are still under debate and vary among 3.60 eV \cite{Mukherjee2022}, 5.09 eV \cite{Kalita2016}, and 7.85 eV \cite{Davidson1972}. We then aimed to tackled this issue by implementing the modified screened hybrid density functional of  Heyd--Scuseria--Ernzerhof (HSE) with the Hartree-Fock exact exchange ($\alpha$) ratio of 0.37 to circumvent the common underestimation from the conventional Perdew–Burke-Ernzerhof (PBE) \cite{Freysoldt2014}. For the sake of computation time, we initially relaxed the structure using the common PBE; then, we applied the HSE functional to estimate the bandgap. This in turn yields 7.91 eV, which agrees with the experimental value \cite{Davidson1972} and another DFT work \cite{Vatti2016}.
\section*{Data availability}
All data from this work is available from the authors upon reasonable request.

\section*{Notes}
The authors declare no competing financial interest.

\begin{acknowledgements}
This work was funded by the Federal Ministry of Education and Research (BMBF) under grant number 13N16292 and the Deutsche Forschungsgemeinschaft (DFG, German Research Foundation) - Projektnummer 445275953. The authors acknowledge support by the German Space Agency DLR with funds provided by the Federal Ministry for Economic Affairs and Climate Action BMWK under grant numbers 50WM2165 (QUICK3) and 50RP2200 (QuVeKS). T.V. and V.D. acknowledge support from the DFG via the SFB 1375 NOA Project C2 (Project number 398816777). This research is part of the Munich Quantum Valley, which is supported by the Bavarian state government with funds from the Hightech Agenda Bayern Plus. The major instrumentation used in this work was funded by the Free State of Thuringia via the projects 2015 FOR 0005 (ACP-FIB) and 2017 IZN 0012 (InQuoSens). C.C. acknowledges a Development and Promotion of Science and Technology Talents Project (DPST) scholarship by the Royal Thai Government. The computational experiments were supported by resources of the Friedrich Schiller University Jena supported in part by DFG grants INST 275/334-1 FUGG and INST 275/363-1 FUGG. S.S. acknowledges funding support by Mahidol University (Fundamental Fund: fiscal year 2023 by National Science Research and Innovation Fund (NSRF)) and from the NSRF via the Program Management Unit for Human Resources \& Institutional Development, Research and Innovation (grant number B05F650024). T.S. is supported via the BMBF funding program Photonics Research Germany (‘LPI‐BT1‐FSU’, FKZ 13N15466) and is integrated into the Leibniz Center for Photonics in Infection Research (LPI). The LPI initiated by Leibniz‐IPHT, Leibniz‐HKI, UKJ and FSU Jena is part of the BMBF national roadmap for research infrastructures.
\end{acknowledgements}
%\clearpage

\section{Supplementary Information}
(1) Thickness measurements of Mica Flake, (2) Electron Irradiation Process, (3) Power-resolved photoluminescence measurements, (4) Spectrum measurements from diffraction-limited spots in mica, (5) Additional data on sample hBN-2, (6) Optical characterization of SiC and GaN, (7) Electronic band structures of various defects in mica.

\bibliography{main_Mica}

%\begin{figure*}
%    \includegraphics{toc/toc_v3_10.png}
%\end{figure*}
 
\end{document}